\documentstyle[sprocl,epsf]{article}
\def\Journal#1#2#3#4{{#1} {\bf #2}, #3 (#4)}
\def\NPB{{\em Nucl. Phys.} B}
\def\PLB{{\em Phys. Lett.}  B}

\def\PRD{{\em Phys. Rev.} D}

\def\be{\begin{equation}}
\def\ee{\end{equation}}
\def\bea{\begin{eqnarray}}
\def\eea{\end{eqnarray}}
\def\b{\begin{eqnarray*}}
\def\e{\end{eqnarray*}}

\def \( {\left(}
\def \) {\right)}

\def\[{\left[}
\def\]{\right]}
\def\lsim {~^{<~}_{\sim~}}
\def\gsim {~^{>~}_{\sim~}}

\def\f{\frac}
\def\tr{\makebox{Tr}}

\def\Gmumu{G_{\mu\mu}}
\def\Amu{A_{\mu}}
\def\Umu{U_{\mu}}

\def\Meff{M_{\makebox{eff}}}
\begin{document}
\title{Gluon Propagator in Maximally Abelian Gauge and Abelian Dominance for
Long-Range Interaction}
\author{K.Amemiya and H.Suganuma}
\address{Research Center for Nuclear Physics (RCNP), Osaka University\\
Mihogaoka 10-1, Ibaraki, Osaka 567, Japan \\
E-mail: amemiya@rcnp.osaka-u.ac.jp} 

\maketitle\abstracts{ We study the gluon propagator in the MA gauge with
the lattice QCD Monte Carlo simulation. The simulation is performed using
the heat-bath algorithm on the SU(2) lattice with $12^3 \times 24$ and
$\beta = 2.3, 2.35$. The propagator of the off-diagonal charged gluon 
behaves as the massive gauge boson and provides the short-range interaction,
while the diagonal gluon propagates long distance.
This is the origin of the abelian dominance in the long-range physics. }
\section{Introduction}
The analogy between the QCD vacuum and the superconductor is 
an interesting current topics for the study of the confinement
mechanism \cite{YN}.

The key point is the use of the maximally abelian(MA) gauge,
because QCD reduces into the abelian gauge theory including
QCD-monopoles\cite{GtH81}.
Remarkably, in the MA gauge, the diagonal part of gluon field plays 
a dominant role to the nonperturbative quantities like confinement and 
chiral symmetry breaking\cite{SST}. 
On the other hand, the off-diagonal part of gluon field behaves as 
a charged matter field and 
does not contribute to the long-range phenomena. 
This is called as the abelian dominance\cite{EI}.
The abelian dominance is confirmed by the recent lattice QCD simulation
\cite{SH91}.
However, the origin of the abelian dominance in the MA gauge 
is not understood yet. 

As a possible physical interpretation for the abelian dominance, 
the effective mass of the charged gluon may be induced in the MA gauge,
and therefore the charged gluon propagation is limited within the short-range 
region because the massive particle propagates within the inverse of its mass.
Here, we study the gluon propagator\cite{ANaka} in the MA gauge
in terms of the interaction range and strength
using the lattice QCD Monte-Carlo simulation. 
\section{Maximally Abelian (MA) Gauge}
In the lattice QCD, the MA gauge is defined by maximizing \\
$R_{\small MA} \equiv \sum_{s,\mu}  
\tr \[ \Umu(s) \tau^3 \Umu^{\dagger}(s) \tau^3 \]$
using the SU(2) gauge transformation.
In the MA gauge, the SU(2) link variable $\Umu(s)$ becomes U(1)-like
due to the suppression of the off-diagonal component.
As for the residual U(1) gauge symmetry, 
we impose the U(1) Landau gauge fixing 
to extract most continuous gauge configration and 
to compare with the continuum theory.
\begin{figure}
\epsfbox{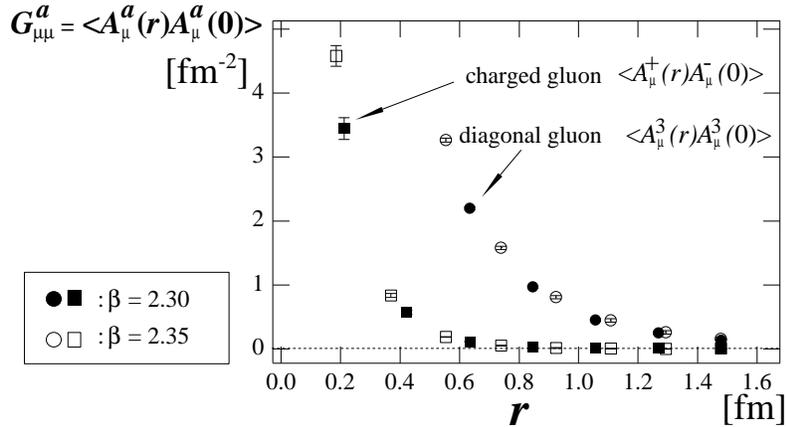}
\caption{The gluon propagator in the MA gauge. In the MA gauge, the only diagonal-gluon is dominant in the long-range region, $r \gsim 0.4$ fm.}
\end{figure}

\section{Lattice QCD Results for Gluon Propagator}
We calculate the gluon propagator in the MA gauge by the lattice QCD 
Monte Carlo simulation, particularly considering 
the scalar combination of 
$\Gmumu^a(r)\equiv \sum^4_{\mu=1}\langle \Amu^{~a}(r)\Amu^{~a}(0)\rangle~(a=1,2,3)$.Here, the scalar combination $\Gmumu^a(r)$ is useful to observe 
the interaction range of the gluon,
because it depends only on the four-dimensional Euclidean radial coordinate 
$r \equiv (x_\mu x_\mu)^{\f{1}{2}}~$.

In Fig.1, in the MA gauge, 
the off-diagonal (charged) gluon propagates within the short-range region 
$r~^{<~}_{\sim~}0.4$ fm : 
the charged gluon behaves as the massive particle 
and does not contribute to the long-range physics in the MA gauge.
On the other hand, the diagonal gluon propagates over the long distance
and influences the long-range physics.
Thus, we find the abelian dominance for gluon propagator 
that the only diagonal gluon field is relevant for the long-range physics
in the MA gauge.
This is the origin of the abelian dominance for the long-range physics.

The off-diagonal (charged) gluon propagator decreases more strongly than 
the massless gauge-boson propagator 
$\Gmumu(r)=\f{3}{4\pi^2}\f{1}{r^2}$.
Therefore, the charged gluon is expected to have an effective mass 
in the MA gauge. We estimate the effective mass of the charged gluon 
from  the scalar combination $\Gmumu(r)$ of the gluon propagator.
Since the propagator of the massive gauge boson with mass $M$ behaves as 
the Yukawa-type function $\Gmumu(r)=\f{3}{4\pi^2}\f{1}{r^2}\exp(-M~r)$,
we can estimate the effective mass of gluons $\Meff$ from the slope of the 
logalithmic plot of $r^2\Gmumu(r)\sim \exp(-M_{\mbox{\small eff}} ~r)$ .
In Fig.2,
the charged gluon correlation $r^2\Gmumu(r) $ decreases linearly in the 
long-range region $r\gsim 0.4$ fm.
We obtain the effective mass of the charged gluon from this slope in the 
intermediate region $r=0.35 \sim 1.0$ fm as 
$M_{\mbox{\small eff}} \approx 4.5 ~\mbox{fm}^{-1}~=0.9~\mbox{GeV}$ .

To summarize, the effective mass of the off-diagonal (charged) gluon
is induced as $\Meff \simeq 1$ GeV in the infrared region in the MA gauge.
Accordingly, the off-diagonal gluon can be neglected and does not contribute to
the long-range physics as $r \gsim 0.4$ fm, 
although its effect appears in the short distance as $r\lsim 0.4$ fm.
Thus, only the diagonal gluon propagates the long distance,
which leads to the origin of the abelian dominance for 
nonperturbative QCD.
\begin{figure}
\begin{minipage}[t]{4.7in}
~~~~~
\epsfbox{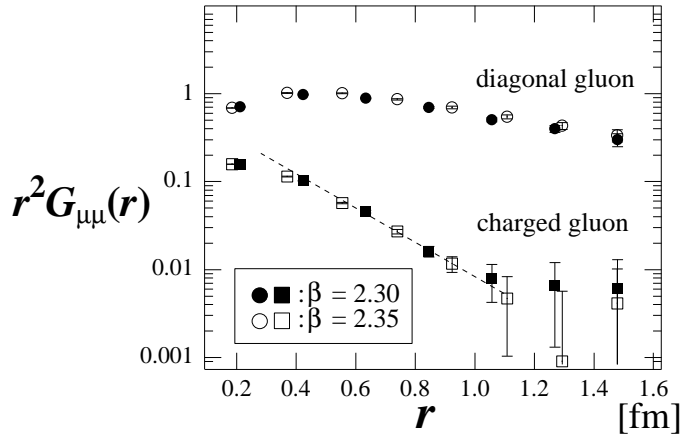}

\caption{The logalithmic plot of $r^2\Gmumu(r)$ as the function 
of the distance $r$ in the MA gauge. The charged gluon propagator behaves
as the Yukawa-type function, $\Gmumu(r)\sim \f{\exp(-Mr)}{r^2}$.
The effective mass of the charged gluon can be estimated
by the slope of the dotted line. }
\end{minipage}
\end{figure}

\end{document}